\documentclass{aa} 
\usepackage{graphicx}
\usepackage{txfonts}

\begin{document}
\title{Intermediate-velocity gas observed towards the Shajn 147 SNR}

\author{S. Sallmen\inst{1} \and B.Y. Welsh\inst{2} }

\offprints{S. Sallmen, \email{sallmen.shau@uwlax.edu } }

\institute{Department of Physics, Univ. of Wisconsin - La Crosse, La Crosse, WI 54601 \and Experimental Astrophysics Group, Space Sciences Laboratory, UC Berkeley, Berkeley}

\date{Received 16 January 2004 / Accepted 12 July 2004}

\abstract{We present high-resolution spectra (R $\sim$ 3 km~s$^{-1}$) 
of the 
interstellar \ion{Na}{i} and \ion{Ca}{ii} interstellar absorption lines observed
towards 3 early-type stars with
distances of 360 to 1380pc along the 
line-of-sight towards the 800pc distant Shajn 147 (S147) Supernova
Remnant (SNR). These data  are supplemented with
far-UV (912 - 1180\AA) aborption spectra 
of \object{HD 36\,665} and \object{HD 37\,318} recorded with the NASA
Far Ultraviolet Spectroscopic Explorer ($\it FUSE$) satellite.
The observations reveal intermediate-velocity (IV) absorption
features at
V$_{helio}$ = +92 km~s$^{-1}$ towards \object{HD 37\,318} and
at V$_{helio}$ = -65 
$\&$ -52 km~s$^{-1}$
towards \object{HD 36\,665}, in addition to several other gas cloud components
with lower velocity. These IV components can be associated
with the expansion of the SNR that has
disrupted the surrounding interstellar
gas.

The IV component at V = +92 km~s$^{-1}$ seen towards \object{HD 37\,318} was
detected only in the far-UV lines of \ion{Fe}{ii} and \ion{N}{ii}, suggesting
that it is composed mainly of warm and ionized gas. 
The two IV components observed towards
\object{HD 36\,665} were detected in \ion{Na}{i}, \ion{Ca}{ii}, \ion{N}{i}, \ion{N}{ii}, \ion{O}{i} and \ion{Fe}{ii}, indicating
that it is composed of both neutral and ionized gas shells.
Highly ionized gas was detected in the \ion{O}{vi}($\lambda$ 1032\AA) 
absorption line at V $\sim$ +40 km~s$^{-1}$ towards
both stars. This hot and highly ionized gas component is
characterized by a
columnn density ratio of N(\ion{C}{iv})/N(\ion{O}{vi}) $<$ 0.27, which is
consistent with that predicted by current models of evolved SNRs.
However, we cannot preclude its origin in the interstellar medium
in line-of-sight to S147.

Column-density ratios of [Mg/Fe], [Al/Si],[Si/Fe], [N/Fe],
[O/Fe] and [Na/Ca]
have been
derived for the IV gas components detected towards S147.
Similar ratios have also been derived for fast-moving gas
observed towards two other SNRs in 
order to gain some insight into the behavior of
element abundances in the disturbed interstellar
gas associated with these regions. 
In all cases except for Na and Ca, these elements
appear to be present with near-solar abundance ratios. 
}

\titlerunning{High-velocity gas towards S147}
\authorrunning{S. Sallmen $\&$ B.Y.Welsh}
\maketitle

\section{Introduction}
Observations of supernova remnants (SNRs)
can provide important information concerning both
the energy and ionization balance of the
interstellar medium (ISM), since supernova explosions
and the strong stellar winds from OB stars are the dominant
sources of high-temperature, high-velocity shocked
gas in the ISM. The majority of the immense energy that is
released in a supernova event is transfered into the
kinetic energy of the gas
and dust ejecta, which then subsequently
interacts and changes both the physical
and chemical state of the ambient interstellar
gas. The spectral (emission and absorption) signatures
of the disturbed low-density gas surrounding SNRs are
best sampled in the ultraviolet regime, which contains
spectral lines covering a wide range of ionization stages
from a large variety of elements. When coupled with
high spectral resolution visible observations (which
provide a key diagnostic of the kinetics of the disturbed
ambient interstellar gas), it is possible to present
a detailed picture of the ionization state, chemical abundance
and velocity structure of the remnant's interaction
with the surrounding ISM.

\begin{figure*}
{\includegraphics[width=12cm]{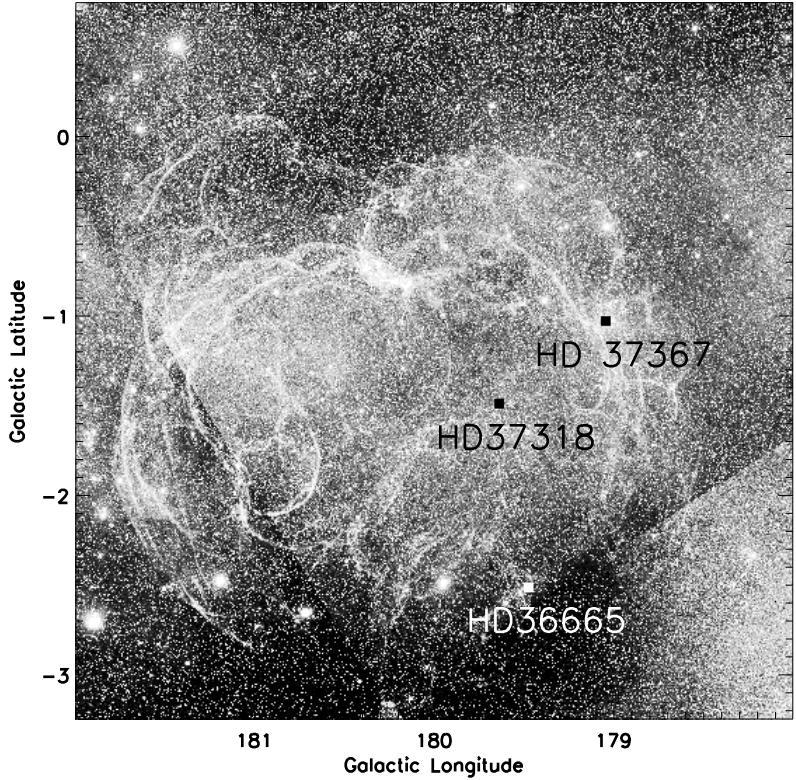}}
\caption{Positions of the 3 stars (1) \object{HD 36\,665}, (2) \object{HD 37\,318} and (3) \object{HD 37\,367} with respect to the H-$\alpha$ emission from the Shajn 147 SNR. The digitized Sky 
Survey image was taken 
from the SkyView web-site at http://skyview.gsfc.nasa.gov.}
\label{Figure 1}
\end{figure*}

Recently we have been carrying out a program of absorption
observations of nearby galactic SNR sight-lines using both the
NASA Far Ultraviolet Spectroscopic Explorer ($\it FUSE$) satellite
(Moos et al. \cite{moos00})
and ground-based visible observations of the
interstellar \ion{Na}{i} and \ion{Ca}{ii} absorption lines.
Thus far we have analyzed the absorption characteristics
of the disturbed interstellar gas associated with the
Monoceros Loop SNR (Welsh et al. \cite{welsh01}, \cite{welsh02a}), the
Cygnus Loop SNR (Welsh et al. \cite{welsh02b}),
the RCW 114 nebula (Welsh et al. \cite{welsh03a}) and the
IC 443 remnant (Welsh \& Sallmen \cite{welsh03b}).
These observations have generally revealed
a complex pattern of shocked neutral and ionized 
shells
of interstellar gas surrounding
these remnants, with gas cloud velocities  
as high as $\sim$ -100 km~s$^{-1}$ being detected.
An overabundance of the refractory elements of Fe, Si and Al 
has been found for all these disturbed interstellar regions,
consistent with the notion that substantial dust grain sputtering has
taken place within the expanding shells of gas.
In contrast, the abundances of N, O and Ar derived from their 
lowest ionization stage are less than the total abundances of the 
elements found in the general ISM, suggesting that
ionization processes may be dominating 
the depletions inferred for the fast-moving gas (Jenkins et al.
\cite{jenkins98}).
Absorption associated with the high-ionization,
high-temperature \ion{O}{vi}($\lambda$1032\AA) ion has not been detected
in the gas surrounding old ($>$ 30,000 years) SNRs.
However, towards the
far younger Vela SNR high-velocity (V$>$ 100 km~s$^{-1}$)
absorption features have been detected in the high-ionization
lines of \ion{C}{iv}($\lambda$1550\AA),
\ion{Si}{iv}($\lambda$1402\AA) and \ion{O}{vi} (Jenkins et al. \cite{jenkins84}). 
In addition, $\it emission$ observations of these 
high-ionization UV lines,
seen in the fast
shocked gas filaments of SNRs, have placed important constraints
on the dynamical pressure and shock mechanisms present in these
highly disturbed regions
of the ISM (Sankrit et al. \cite{sank03},
Sankrit \& Blair \cite{sank02}).  

\begin{table*}
\caption{Stellar Target Information}
\label{table:1}
\centering
\begin{tabular}{c c c c c c c c}
\hline
\hline
Star & (l,b) & m$_{v}$ & Sp & E(B-V)& V sin i& R.V.& distance \\
&&&&& km s$^{-1}$&km s$^{-1}$&(pc) \\
\hline
\object{HD 36\,665}&(179.5$^{\circ}$,-2.5$^{\circ}$)&8.2&B0Ve&0.64&60&+9.0& 880 \\
\object{HD 37\,318}&(179.6$^{\circ}$,-1.5$^{\circ}$)&8.4& B1IV& 0.61&50&+5.8& 1380 \\
\object{HD 37\,367}&(179.0$^{\circ}$,-1.0$^{\circ}$)&6.0&B2IV& 0.40&20&+29.6& 361(+150,-85)* \\
\hline
\multicolumn{6}{l}{\sl * = Hipparcos distance} \\
\hline
\hline
\end{tabular}
\end{table*}

The Shajn 147 SNR ($\it l$ = 180.3$^{\circ}$, $\it b$ = -1.7$^{\circ}$) is 
an optically faint, highly filamentary galactic supernova
remnant (SNR) of age $\sim$ 10$^{5}$ years with a distance
of $\sim$ 800pc (Kundu et al. \cite{kundu80}).
It has a shell-like structure of diameter 3$^{\circ}$ as revealed
by both radio (Furst \& Reich \cite{furst86}) and visible
(Minkowski \cite{mink58}) observations. No X-ray emission has been
detected from this remnant (Sauvageot et al. \cite{sauv90}).
Optical absorption studies of the interstellar \ion{Ca}{ii} K-line at 3933\AA\
observed towards the star \object{HD 36\,665} (d $\sim$ 880pc)
by Silk \& Wallerstein (\cite{silk73}) revealed
intermediate-velocity (IV)
gas with a velocity of $\sim$ - 69 km~s$^{-1}$ that was associated
with the expansion of the SNR blast wave into the surrounding ISM.
Follow-up $\it I.U.E.$ absorption studies of the sight-lines
towards the stars \object{HD 36\,665} and \object{HD 37\,318} confirmed the
disturbed state of the ISM surrounding Shajn 147, with absorption 
features being detected at
respective velocities of V$_{helio}$ = - 63 km~s$^{-1}$
and V$_{helio}$ = +92 km~s$^{-1}$ (Phillips, Gondhalekar \& Blades
\cite{phillips81} hereafter PGB81; Phillips \& Gondhalekar
\cite{phillips83} herefafter PG83). 
These IV absorption components were detected in UV
lines such as \ion{C}{ii}, \ion{C}{ii}*, \ion{Al}{ii} and \ion{Fe}{ii}, but not in the
high-ionization (T $\sim$ 80,000K) lines 
of \ion{C}{iv} or \ion{Si}{iv}. Similarly, no IV absorption features were detected in
neutral ions such as \ion{C}{i} or \ion{N}{i} towards either star.
From these UV data it was concluded that
the shocked gas surrounding the SNR is not strongly ionized,
consistent with what may be expected from an evolved remnant.

In this Paper we present high spectral resolution (R $\sim$ 3 km~s$^{-1}$)
observations
of the visible interstellar \ion{Na}{i} and \ion{Ca}{ii} absorption lines seen towards
3 stars (\object{HD 36\,665},
\object{HD 37\,318} and \object{HD 37\,367}) with sight-lines towards Shajn 147.
These data are complemented by
far-ultraviolet absorption observations towards
the former two stars (that have distances comparable with Shajn 147) 
recorded over the 912 - 1180\AA\ range 
using the NASA $\it FUSE$ satellite. Our observations confirm the
presence of a complex of absorption components that span a range of
velocities from -65 to +90 km~s$^{-1}$, with the highest-velocity
components being associated with the expansion of the SNR into
the ambient interstellar medium. Our far-UV observations suggest
that these intermediate-velocity clouds contain both neutral
and ionized gas and that the refractory elements present in these
components possess near-solar abundances. 

\section{Observations}
We have carried out visible
absorption observations towards \object{HD 36\,665}, 
\object{HD 37\,318} and the
foreground star \object{HD 37\,367} using the Aurelie echelle spectrograph at the 1.52m
telescope of the Observatoire de Haute Provence (France)
on the nights of September 26 - 29th, 2002. Data were
recorded separately at the \ion{Na}{i}($\lambda$5890\AA)
and \ion{Ca}{ii}($\lambda$3933\AA)
interstellar lines.  
In Table \ref{table:1} we list the relevant astronomical data for the 3
stars, which includes their
visual magnitudes, spectral
types, reddening values, rotational velocity (V sin i),
radial velocity (R.V.) and distance estimates. These
values were taken from
the on-line Simbad astronomical data-base, with the distance
and reddening
estimates being taken from PGB81,  
Savage et al. (\cite{savage85}) and
the Hipparcos catalog (ESA \cite{esa97}). For \object{HD 36\,665} and \object{HD 37\,318}
we derived averaged rotational velocities from the widths of
several prominent stellar lines in their visible and far-UV spectra.
The spectral type for \object{HD 36\,665} was taken from Halbedel et al. (\cite{halb96}).
The
positions of the 3 targets with respect to the
optical H-$\alpha$ emission contours of the Shajn 147
SNR are shown in Figure \ref{Figure 1}.

The photon data
were recorded with an EEV 2048 x 1024 CCD detector
and the raw spectral images extracted using data
reduction procedures analogous to those
detailed in Sfeir et al. (\cite{sfeir99}). Briefly
these software programs perform cosmic ray removal,
background subtraction, flat-fielding,
optimal spectral order extraction and wavelength assignment
(from Th-Ar calibration spectra) on the recorded data.
The resolution of the resultant spectra was 2.75 km~s$^{-1}$
and the wavelength accuracy of the calibrated data was $\sim$ $\pm$0.02 \AA.
The S/N ratio (per pixel) of the recorded spectra were $\sim$ 30:1 for the \ion{Na}{i}
spectra and $\sim$ 20:1 for the \ion{Ca}{ii} data.
All velocities subsequently quoted in this Paper
are reported in the heliocentric frame
of reference. Finally, we have removed the many telluric
water vapor absorption lines that particularly affect contamination
of the \ion{Na}{i} D-line-profiles using a computed synthetic atmospheric
scaled transmission spectrum as described in Lallement et
al. (\cite{lallement93}).

The $\it FUSE$ spectra of \object{HD 36\,665} and \object{HD 37\,318} were taken using the
LWRS (30 x 30 arc second) spectrograph aperture, with the data being
recorded in the detector time-tag mode. Photon events were recorded
for a total exposure time of 26 ksec for \object{HD 36\,665} and 35 ksec
for \object{HD 37\,318}. The data were processed using version
CFv2.2.3 of the $\it FUSE$ science data reduction (CALFUSE) pipeline,
which corrects for geometrical image distortions, background
subtraction, image thermal drifts, detector deadtime and
wavelength calibration (Sahnow et al. \cite{sahnow00}). Due to the
relatively high reddening of both objects, the far-UV spectra at
wavelengths $<$ 1000\AA\ were of a low S/N ratio ($<$ 5:1),
whereas the spectra were better exposed at longer wavelengths.
Therefore, only the data contained in the LiF1a and  LiF1b 
spectral channels (i.e. $\lambda$ $>$ 1000\AA)
have been utilized in this analysis.
Since the $\it FUSE$ instrument does not provide absolute wavelength
accuracies to better than $\pm 20$ km~s$^{-1}$, we have
determined the far-UV wavelength scale with
reference to the many H$_{2}$ molecular lines that were detected
in these UV spectra. We have assumed that these lines are formed
at the same (heliocentric) velocity as that of the main absorption
component detected in the \ion{Na}{i}
interstellar D-lines at V$_{helio}$ =
+ 14.5 km~s$^{-1}$ for \object{HD 36\,665} and
at V$_{helio}$ = + 12.0 km~s$^{-1}$ for \object{HD 37\,318} (see Table \ref{table:2}). 
We estimate that the accuracy of our derived far-UV wavelengths
is $\sim$ $\pm5$ km~s$^{-1}$.  
The resolution of the LiF channel data used in the subsequent analysis
was
found to be R $\sim$ 21,000 (i.e. 14 km~s$^{-1}$),
as determined from fitting several weak interstellar absorption lines
in the 1020 - 1180\AA\ wavelength region. 

\section{Interstellar Analysis}
\begin{table*}[htbp]
\caption{\ion{Na}{i} and \ion{Ca}{ii} Absorption Line Best-Fit Parameters (stars listed by
increasing distance)}
\label{table:2}
\centering
\begin{tabular}{ccccccccccc}
\hline
\hline
Star&V&$\it b$&N&&&V&$\it b$&N& \ion{Na}{i}/\ion{Ca}{ii} \\
&km s$^{-1}$&&(10$^{10}$ cm$^{-2}$)&&&km s$^{-1}$&&(10$^{10}$ cm$^{-2}$)& \\
\hline
{\bf \object{HD 37\,367}} &&&& \\
...\ion{Na}{i}...&&&&&...\ion{Ca}{ii}...&+1.3&4.4&19$\pm3$&- \\
&+13.5&3.6&620$\pm30$&&&+16.7&6.0&125$\pm30$&5.0 \\
&&&&&&+22.0&3.8&85$\pm20$&- \\
&&&&&&+31.5&7.9&41$\pm15$&- \\
{\bf \object{HD 36\,665}} &&&& \\
...\ion{Na}{i}...&-67.0&4.1&7.6$\pm0.5$&&...\ion{Ca}{ii}...&-66&3.8&90$\pm30$& 0.084 \\
&&&&&&-56.0&7.5&72$\pm23$& - \\
&&&&&&-48.0&8.0&44$\pm12$& - \\
&&&&&&-34.0&2.7&14$\pm4$& - \\
&&&&&&-6.0&4.8&55$\pm12$&- \\
&+2.8&3.6&190$\pm35$&&&+6.0&4.9&150$\pm25$& -\\
&+14.5&4.7&$>$1350*&&&+17.5&7.0&380$\pm40$&$>$3.6  \\
&+34.0&0.8&6.9$\pm1$&&&+30.0&3.0&55$\pm20$&0.13\\
&+39.5&2.1&2.2$\pm0.5$&&&+36.0&4.8&45.0$\pm8$&0.049\\
{\bf \object{HD 37\,318}} &&&& \\
...\ion{Na}{i}...&-35.5&4.0&4.6$\pm1.0$&&...\ion{Ca}{ii}...&-33.0&2.2&6.6$\pm2.0$& 0.7 \\
&+12.0&5.7&$>$2500*&&&+13.0&6.4&205$\pm40$&$>$120 \\
&+24.5&2.6&55$\pm10$&&&+21.0&7.4&290$\pm50$& - \\
&+37.0&3.8&7.9$\pm1.5$&&&&&$<$3.0& $<$2.6 \\
\hline
\multicolumn{9}{l}{\sl * = saturated component} \\
\hline
\hline
\end{tabular}
\end{table*}

\begin{table*}[htbp]
\caption{Best-Fit Parameters to the IV components of the far-UV absorption lines}
\label{table:3}
\centering
\begin{tabular}{ccccccccc}
\hline
\hline
Line&V&$\it b$&N&&V&$\it b$&N \\
&km s$^{-1}$&&(10$^{12}$cm$^{-2}$)&&km s$^{-1}$&&(10$^{12}$cm$^{-2}$) \\
\hline
{\bf \object{HD 36\,665}}&&&&&&&& \\
...\ion{Fe}{ii} 1144.9\AA...&-65&9.0&33$\pm6$&&-48&9.2&23$\pm8$ \\
...\ion{Fe}{ii} 1122.0\AA...&-64&9.0&33$\pm6$&&-55&9.2&23$\pm8$ \\
...\ion{N}{i} 1135.0\AA...&-64&6.7&76$\pm20$&&-51&5.4&24$\pm5$ \\
...\ion{N}{i} 1134.2\AA...&-63&6.7&76$\pm20$&&-49&5.4&24$\pm5$ \\
...\ion{N}{ii} 1084.0\AA...&&&&&-53&18.0&240$\pm30$* \\
...\ion{O}{i} 1039.2\AA...&-65&5.4&270$\pm50$&&-56&4.2&330$\pm80$ \\
...\ion{O}{vi} 1031.9\AA...&+10&17.5&185$\pm40$&&+45&9.0&36$\pm5$ \\
&&&&&&&& \\
{\bf \object{HD 37\,318}}&&&&&&&& \\
...\ion{Fe}{ii} 1144.9\AA...&&&&&+92&3.7&40$\pm7$ \\
...\ion{Fe}{ii} 1122.0\AA...&&&&&+90&4.2&40$\pm7$ \\
...\ion{N}{ii} 1084.0\AA...&&&&&+92&16.2&155$\pm20$* \\
...\ion{O}{vi} 1031.9\AA...&+11.0&9.5&55$\pm5$&&+38&19.0&33$\pm10$ \\
\hline
\multicolumn{8}{l}{\sl * = saturated component} \\
\hline
\hline
\end{tabular}
\end{table*}

\begin{figure*}
\centering
{\includegraphics[width=17cm]{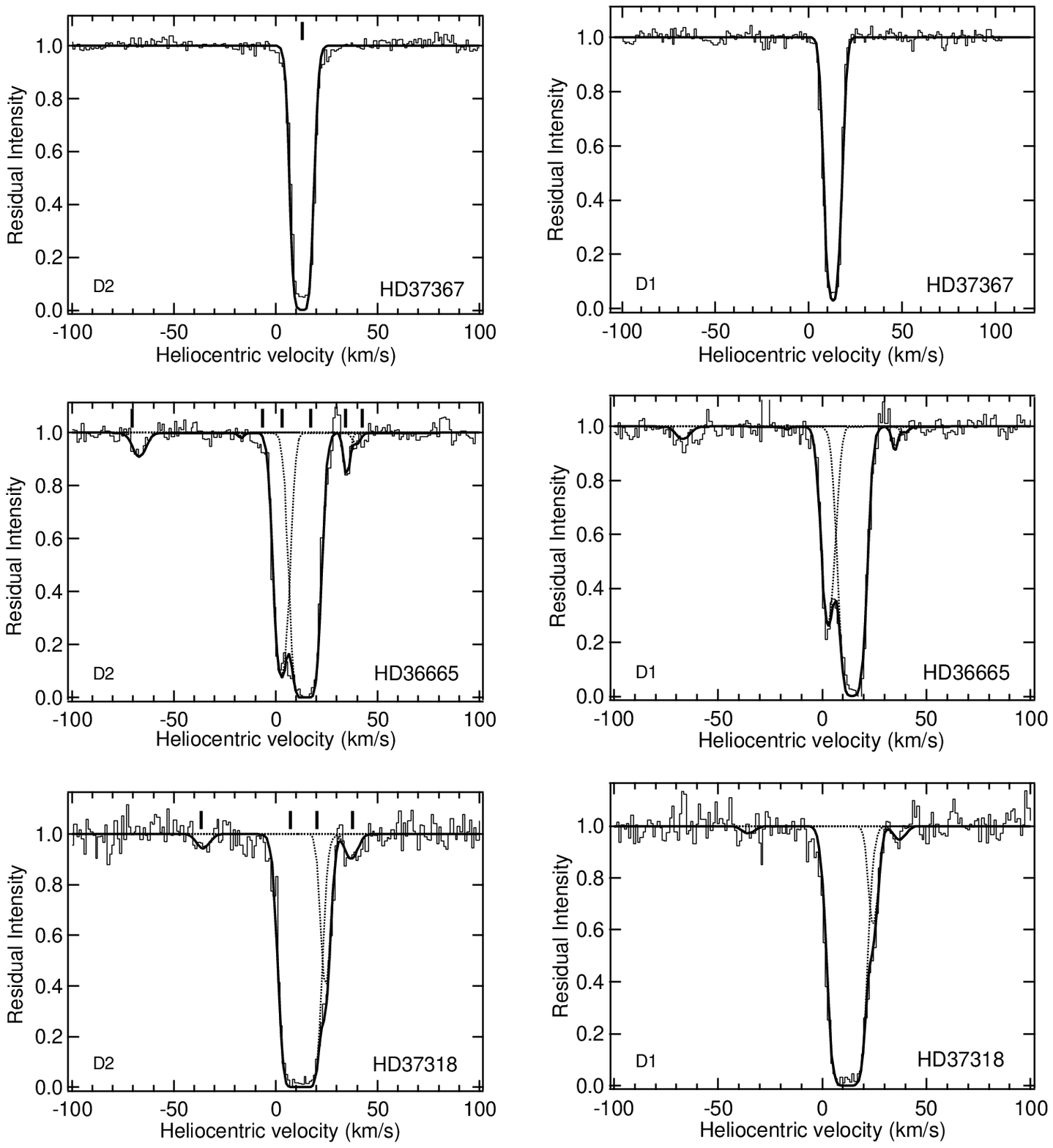}}
\caption{Interstellar \ion{Na}{i} D2 and D1 absorption lines observed towards the 
3 stars in line-of-sight towards the S147 SNR. Superposed on the residual intensity data points (light
full line) is the multi-component best-fit absorption model (see Table \ref{table:2}). 
The dotted lines indicate each of the components used in the model fits.} 
\label{Figure 2}
\end{figure*}

We have determined the local stellar continua for the interstellar
\ion{Na}{i} and \ion{Ca}{ii} absorption lines for the 3 stars using
a multi-order polynomial. Errors associated with this placement
are automatically generated by the software routine and
have been discussed in Welsh et al. (\cite{welsh90}). For
these visible interstellar lines the local continua were
well-behaved and the placement error was small. 
The resultant residual intensity profiles are shown in
Figures \ref{Figure 2} and \ref{Figure 3}.
These profiles were then fit with
multiple absorption
components (identified with interstellar gas `clouds') using a line-fitting
program described in Sfeir et al. (\cite{sfeir99}). This program assigns
a 3-parameter theoretical fit to the observed absorption profiles by
assigning values for the interstellar
gas cloud component velocity, V, a gaussian velocity
dispersion, $\it b$ and a cloud component column density, N.
This fitting procedure works very well for absorption components
that are not saturated, but the process leads to large
uncertainties in the derived column densities for the
central cores of spectral lines that possess
very high values of saturation. In such cases (i.e. for the central
line-core components) the best-fit values of 
$\it b$ and N shown in Table \ref{table:2} (and 
marked with *) are not well constrained
and are listed only for the purpose of showing the velocity range
over which the central absorption occurs.
Fortunately, most of the IV absorption components located far from
the saturated line-cores are unsaturated and
reliable values of V, $\it b$ and N are 
listed in Table \ref{table:2} for both the \ion{Na}{i} and \ion{Ca}{ii} spectral lines.

Our profile fits have been performed using
the $\it minimum$ number of
absorption components, and when
more than one line of a species is available (such as 
the \ion{Na}{i} doublet), both profiles are fit
simultaneously. We have used the
criterion of Vallerga et al. (\cite{vallerga93})
such that the addition of
extra components (which will always
improve the fit at some level)  results in a reduction of 
the chi-squared residual error between the observed and
computed data points of $<$ 1$\%$ .
Errors for the derived component column densities (for
unsaturated components)
are also listed in Table \ref{table:2}.
All of these model fits
are shown
superposed on the observed line-profiles in Figures \ref{Figure 2} and \ref{Figure 3}. 

The far-UV spectra were searched
for interstellar absorption lines
that possesed IV features with velocities consistent with those
found from previous studies of both stars using the
$\it I.U.E.$ satellite (PGB81, PG83).
These IV components (occuring
at V $\sim$ +90 km~s$^{-1}$ for \object{HD 37\,318},
and V $\sim$ -60 km~s$^{-1}$ for \object{HD 36\,665}) were then
fitted and analyzed using the same
techniques as outlined above for the visible interstellar lines.
We note that only for the case of the \ion{O}{vi} ($\lambda$1031.9\AA)
interstellar line was continuum placement judged to be problematic, due to
the presence of nearby \ion{O}{vi} stellar and
molecular H$_{2}$ line features. Our final choice of continuum
placement for the two stars was guided by the examples
given in Lehner et al. (\cite{lehner01}).
It was found that the absorption centered at $\sim$ -60 km~s$^{-1}$,
detected in 5 of the far-UV lines seen towards
\object{HD 36\,665}, was best fit using a two-component model (with
cloud components at V = -64 and -52 km~s$^{-1}$). Absorption
components with a similar velocity-split were also observed
in the \ion{Ca}{ii} spectrum of this sight-line. 
The high-ionization line of \ion{O}{vi} ($\lambda$1031.9\AA)
was strongly detected towards both stellar targets, but
no IV features with velocities in common with those detected
in the lower ionization UV lines were seen. However, a
partially
resolved absorption feature was observed at V $\sim$ +40 km~s$^{-1}$
in the \ion{O}{vi} lines towards both stars and this component was fitted
in tandem with the strong central absorbing component.
Components with a similar velocity ($\sim$ +40 km~s$^{-1}$)
were also detected in the \ion{Na}{i} and \ion{Ca}{ii} lines.
All of the far-UV
absorption lines possessing these
intermediate-velocity features and their associated
model fits are shown in Figures \ref{Figure 4} and \ref{Figure 5}, and the best-fit
UV absorption-line parameters
are listed in Table \ref{table:3}. The best-fit
$\it b$-values for the UV lines are generally large,
indicating the presence of absorption structure that
is unresolved by the $\it FUSE$ instrument. 

We also note that the $\it FUSE$ spectra of both \object{HD 36\,665}
and \object{HD 37\,318} contain many (strong) molecular H$_{2}$ lines 
associated 
with the Lyman (B-X) and Werner (C-X) bands. None of these
lines is accompanied by an
absorption component with a velocity consistent with the IV
features detected by the atomic lines. Similarly,
Gondhalekar \& Phillips (\cite{gond80}) also failed to detect
high-velocity components in the many
ultraviolet $^{12}$C$^{16}$O lines
observed towards \object{HD 36\,665}. 
This is in marked
contrast to the Monoceros Loop SNR in which IV molecular H$_{2}$
was observed at V = +65 km~s$^{-1}$ in six of the J = 3 rotational
levels (Welsh et al. \cite{welsh02a}).

\section{Discussion}
\begin{figure}
\resizebox{\hsize}{!}
{\includegraphics{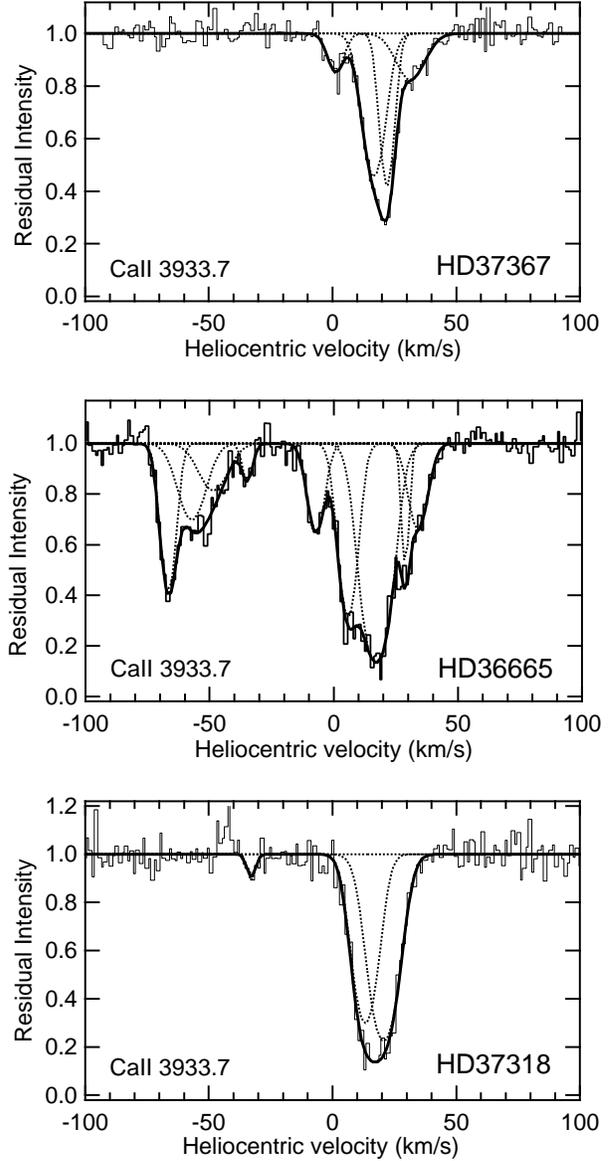}}
\caption{Observed and best-fit models for the interstellar \ion{Ca}{ii}-K absorption profiles recorded towards the 3 target stars. 
Symbols are as stated in Figure \ref{Figure 2}.}
\label{Figure 3}
\end{figure}

It is immediately apparent from Figures \ref{Figure 2}
and \ref{Figure 3} that the \ion{Na}{i} and \ion{Ca}{ii}
absorption profiles recorded towards both \object{HD 36\,665} and \object{HD 37\,318} are
far more complex and span a much wider range in velocity than
the profiles recorded towards \object{HD 37\,367}. Since this latter star lies
in the same sight-line as S147, but is $\sim$ 500pc
foreground to the nominal distance to
the remnant, we can confidently assign the
additional higher-velocity absorption components observed towards
both \object{HD 36\,665} and \object{HD 37\,318} as being due to the interaction 
of the expanding SNR with the surrounding ISM. Our following
discussion will focus on the absorption properties of
these intermediate-velocity (IV) interstellar components.
 
\subsection{The \ion{Na}{i} D-line absorption spectra}
In Figure \ref{Figure 2} we see that
all 3 stars show strong central absorption over the velocity
range of -10 to +30 km~s$^{-1}$, which we associate with
absorption due to the foreground line-of-sight 
interstellar medium towards these targets. From these data
we assign velocities of V = +13.5, V = +14.5 and V = +12.0
km~s$^{-1}$ to the main (line-of-sight) absorption towards
\object{HD 37\,367}, \object{HD 36\,665} and \object{HD 37\,318} respectively.
The absorption
component observed towards \object{HD 36\,665} at V $\sim$ 0 km~s$^{-1}$
is seen to be partially resolved in our spectra. This component
becomes stronger and more saturated in the spectrum of
\object{HD 37\,318}, which is consistent with the greater distance
of the ISM being sampled.

The higher-velocity components observed
in the \ion{Na}{i} spectra at V = -67 and +39 km~s$^{-1}$
towards \object{HD 36\,665} and at V = -35 km~s$^{-1}$ towards \object{HD 37\,318}
presumably originate in the disturbed ISM surrounding the S147 SNR.
These absorption components 
were also detected at similar
velocities in the \ion{Ca}{ii} spectra, and confirm the original
observations of Silk \& Wallerstein (\cite{silk73}) who
first detected IV gas towards \object{HD 36\,665} at V$_{helio}$ = -58 km~s$^{-1}$.
Our observations are also consistent with the H-$\alpha$, \ion{N}{ii}
and \ion{S}{ii} emission line observations of S147 by Kirshner \&
Arnold (\cite{kirsh79}) who derived a $\it systematic$ expansion
velocity of $\sim$ 80 km~s$^{-1}$ for the remnant.

\subsection{The \ion{Ca}{ii} K-line absorption spectra}
The \ion{Ca}{ii} K-line profiles of Figure \ref{Figure 3} clearly show that the strong
foreground line-of-sight absorption is formed over 
a similar velocity range to that
of the \ion{Na}{i} spectra (i.e. 0 to +30 km~s$^{-1}$). 
Additional, higher-velocity components have been detected
at V = -65.5, -55.5, -46.5 and +37.5 km~s$^{-1}$ towards
\object{HD 36\,665}, indicating that this sight-line is clearly complex
and consists of many warmer and/or more ionized fast-moving gas clouds.
The sight-line towards \object{HD 37\,318} is
far less complicated with
the strong central \ion{Ca}{ii} absorption being accompanied by
a single extra low-velocity component at
V = -33.0 km~s$^{-1}$. Figure \ref{Figure 3} clearly shows no sign
of the component detected at V = +90 km~s$^{-1}$ in the
$\it I.U.E.$ data of PG83. Since this IV absorption
component was not detected in either \ion{Na}{i} or \ion{Ca}{ii} towards \object{HD 37\,318},
it must contain mainly ionized (and warmer) gas.

All these data are consistent with 
observations towards other SNRs such as the Monoceros Loop,
the Vela SNR and IC 443, in that the visible absorption profiles
possess a multi-component structure indicative of sight-lines
that pass through a highly disturbed
and/or inhomogenous interstellar medium. The various absorption
components can be identified with expanding gas shells and/or
disrupted ambient clouds of gas. The physical complexity
of these disrupted interstellar clouds can be easily seen
in the many contorted filaments of gas emission shown
in Figure \ref{Figure 1}, which suggests a chaotic pattern of disruption
of the surrounding ISM.
In the following sections we will
attempt to describe the physical and chemical nature of these
disturbed interstellar regions. 

\subsection{The \ion{Na}{i}/\ion{Ca}{ii} ratios}
\begin{figure*}
\centering
{\includegraphics[width=17cm]{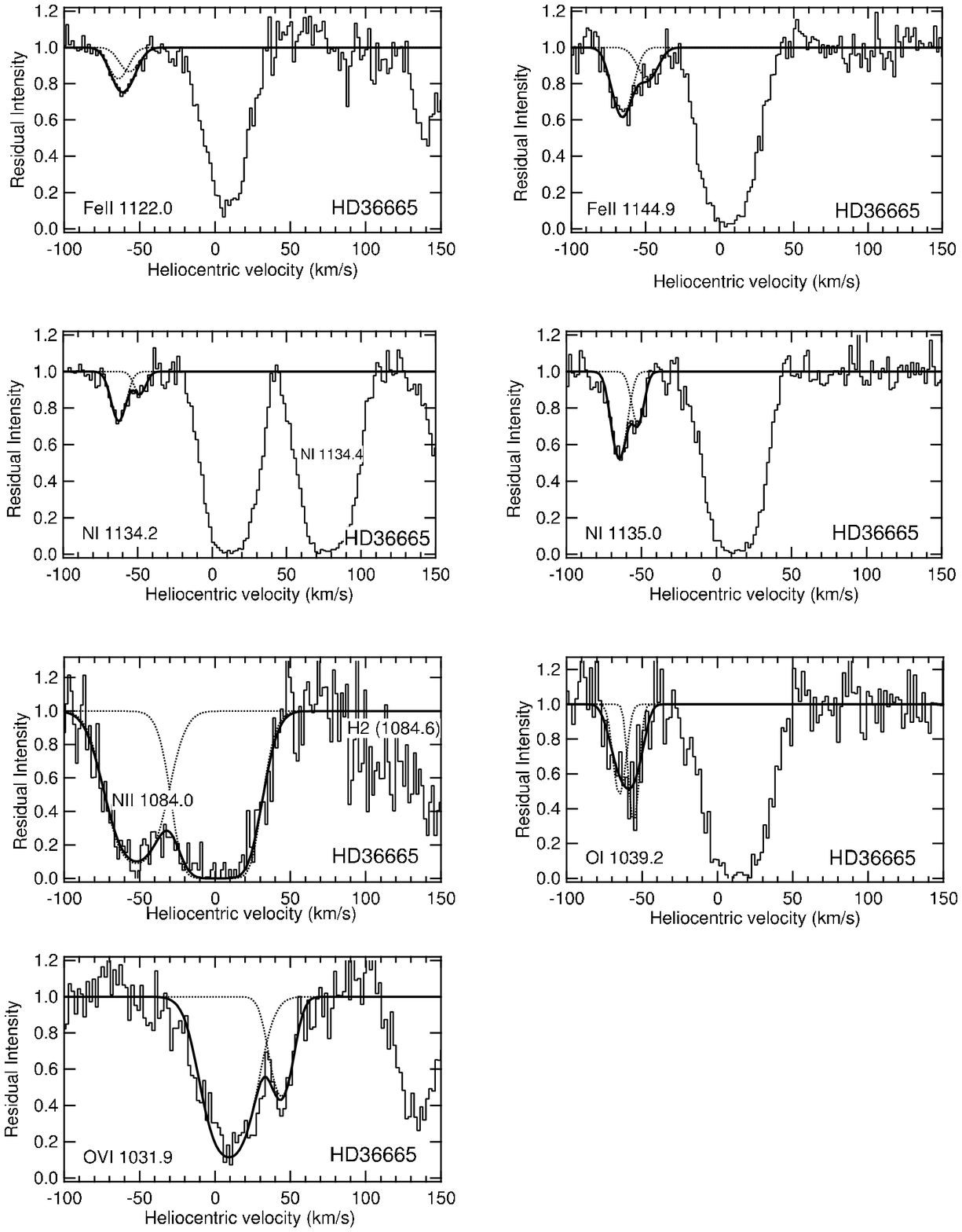}}
\caption{Observed and best-fit models for the high-velocity absorption features recorded in the far-UV lines towards the star HD36\,665}
\label{Figure 4}
\end{figure*}

The column-density ratio of N(\ion{Na}{i})/N(\ion{Ca}{ii}) has been widely
used as a diagnostic of the
physical state of gas clouds in the interstellar
medium (Routly and
Spitzer \cite{routly52}). For low-velocity 
clouds a ratio of $>$ 1.0 is generally found
in the ISM. Our data show similar ratios for
the strong
central foreground absorption components observed towards all 3 stars. 
In contrast, interstellar shocks produced by fast-moving gas clouds
can cause thermal sputtering and collisions of interstellar
dust grains, which then return Ca into the gas phase,  
resulting in a low observed value
($<<$ 1.0) of the \ion{Na}{i}/\ion{Ca}{ii} ratio.
For the (well-resolved) IV components at V = -65 and
+39 km~s$^{-1}$
seen towards \object{HD 36\,665} we obtain 
\ion{Na}{i}/\ion{Ca}{ii} ratios of 0.085
and 0.054 respectively. Similarly low values of this ratio
have been found for the Vela SNR by
Danks \& Sembach (\cite{danks95}),
in which SN-driven shocks were forwarded to
explain the anomalous (i.e. near-solar)
gas-phase abundance ratio of [Na/Ca] in the
high-velocity gas cloud components.
However, we note that
the \ion{Na}{i}/\ion{Ca}{ii} ratio can also reflect the unusual ionization
of Na and Ca due to a nearby strong source of UV photon
flux. We shall comment further on this possibiity in the following
discussion
of the UV absorption properties of the IV gas components observed
towards S147.

\subsection{The far-UV absorption spectra}
\begin{figure*}
\centering
{\includegraphics[width=17cm]{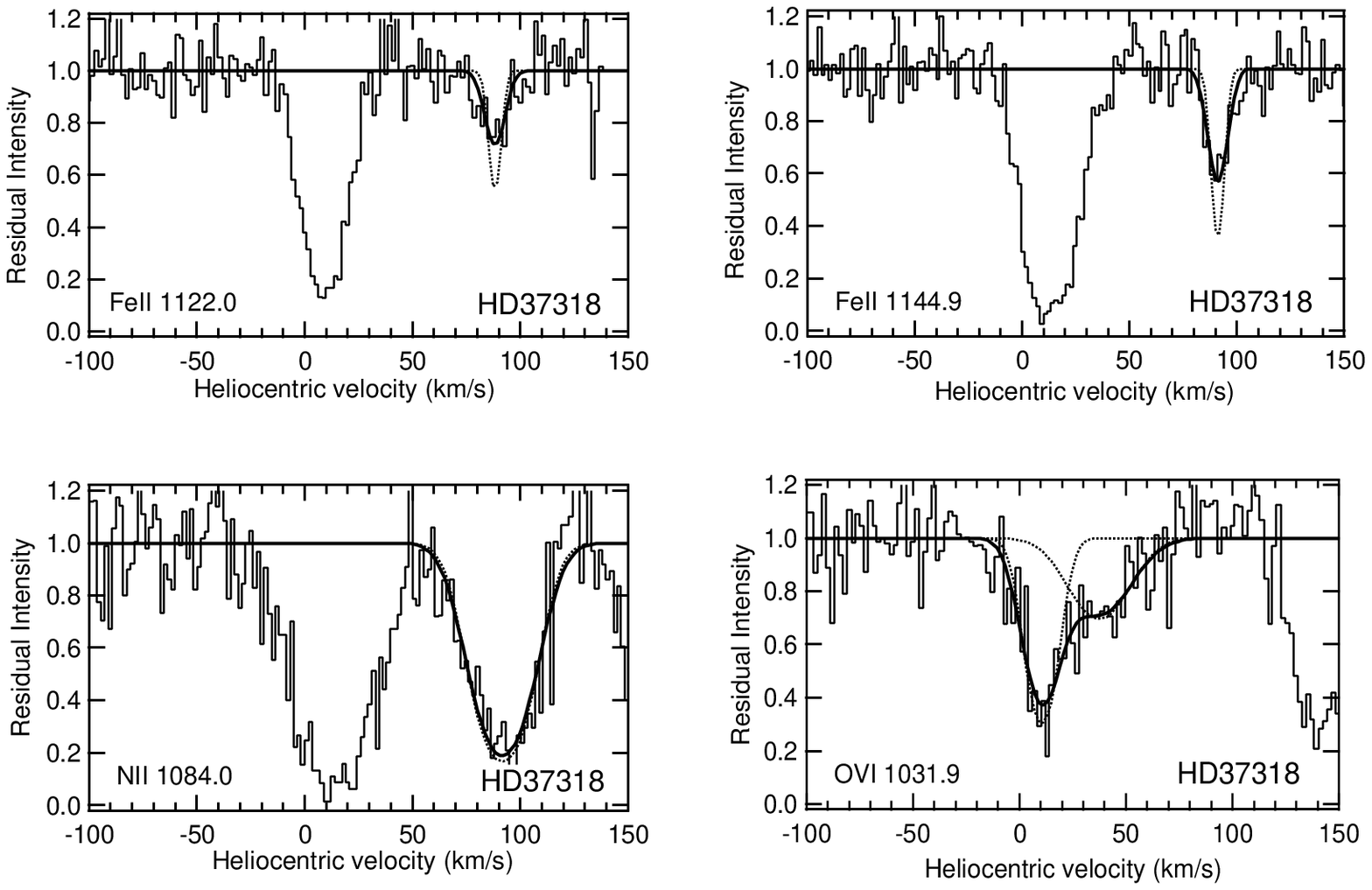}}
\caption{Observed and best-fit models for the high-velocity absorption features recorded in the far-UV lines towards the star HD37\,318.}
\label{Figure 5}
\end{figure*}

As shown in Figure \ref{Figure 4}
our $\it FUSE$ observations have revealed partially resolved 
intermediate-velocity absorption features at V = -64 and -52 km~s$^{-1}$
towards \object{HD 36\,665} in
the UV lines of \ion{Fe}{ii}, \ion{N}{i} and \ion{O}{i}, in addition to a strong
blended component at V = -53 km~s$^{-1}$ in the \ion{N}{ii} line
at 1084.0\AA.
Absorption components with a similar
velocity were also detected 
in our visible absorption spectra towards \object{HD 36\,665},
and by PGB81
in the near-UV lines of \ion{Fe}{ii}, \ion{Mg}{i}, \ion{Mg}{ii}, \ion{Al}{ii}, \ion{C}{ii} and \ion{Si}{ii}.
All these observations
indicate that the gas travelling with these velocities
is not only composed of neutral components (e.g. \ion{Na}{i}, \ion{Mg}{i}
and \ion{O}{i}), but 
it must also contain significant amounts of 
warmer and more ionized components due to the
detection of \ion{Fe}{ii}, \ion{Al}{ii} and \ion{N}{ii}.
Given the very similar velocities of these IV components
it seems likely that the neutral and ionized gas 
may be in the form of co-moving sheets or shells
of interstellar material that have been disrupted by the
expansion of the SNR.

In the case of the sight-line towards \object{HD 37\,318} we note that
the IV feature was
detected only in the far-UV lines of \ion{Fe}{ii} and
\ion{N}{ii} at +92 km~s$^{-1}$. It was present neither in
the \ion{Na}{i} or \ion{Ca}{ii} lines, nor in any of the
low-ionization far-UV lines.
These results are consistent with the near-UV observations of
this sight-line by PG83 
in which this IV component was detected only in the lines of
\ion{C}{ii}, \ion{Mg}{ii}, \ion{Al}{ii}, \ion{Al}{iii}, \ion{Si}{ii}, \ion{S}{ii} and \ion{Fe}{ii}.
All these data
strongly suggest that this component is $\it not$ composed
of appreciable amounts
of cold and neutral gas, but instead is mainly composed 
of (warmer) ionized gas.
We note that the distance to \object{HD 37\,318} ($\sim$ 1380pc) suggests
that the sight-line samples both the near and far sides of the
remnant. This view is strongly supported by the detection
of the +92 km~s$^{-1}$ component which 
is most probably associated with gas receding away
from the SNR. We might also have
expected to detect approaching (negative-velocity) IV gas along
this sight-line. Its non-detection
towards a star of this distance (which lies
well beyond the nominal distance to the SNR)
may well be due to large-scale density 
inhomogeneities in the near-side filamentary shell(s) of the SNR.
Similar behavior
has been observed towards many stars
with sight-lines towards the Vela SNR,
in which maps of the observed high velocity components revealed
a chaotic structure of kinematic motions (due to
interactions with pre-existing
ambient interstellar clouds), rather than a
picture of coherent expansion
of the remnant material (Jenkins et al. \cite{jenkins84}).

Highly ionized \ion{O}{vi}($\lambda$1032\AA)
gas components with a velocity of V $\sim$ +40 km~s$^{-1}$
have been
detected along both stellar sight-lines.
This high-temperature gas cloud could
be foreground to the SNR, but
its anomalously high velocity compared with
gas found in the general ISM suggests that it
is more probably
associated
with the expansion of the SNR.
We note that no IV components were detected
in the high ions of \ion{C}{iv}, \ion{Si}{iv} or \ion{N}{v} towards both stars
by PGB81 and PG83. Using their published upper limit for
intermediate velocity \ion{C}{iv} absorption
towards \object{HD 37\,318} of log N(\ion{C}{iv}) $<$ 12.96 cm$^{-2}$, we obtain
a ratio of N(\ion{C}{iv})/N(\ion{O}{vi}) $<$ 0.27. This ratio only
has physical significance if the \ion{C}{iv} and \ion{O}{vi} coexist in the
same absorbing cloud, and we note that the published upper limit
for \ion{C}{iv} was determined for absorption at V = +90 km~s$^{-1}$
by PG83. However, if we assume that the two species do
physically coexist then this 
ratio is consistent
with that observed in the hot
gas of 
the galactic disk (Spitzer
\cite{spitzer96}). The ratio is also consistent with
that derived
for hot gas in
the evolved supernova bubble model
of Slavin \& Cox (\cite{slavin92}) in which an isolated
SN event explodes into interstellar gas with an ambient
density of 0.1 - 0.3 cm$^{-3}$ and a magnetic field of
1 - 3$\mu$G.  Therefore, at present we cannot confirm
where the V = +40 km~s$^{-1}$ \ion{O}{vi} absorption is taking place.
However we do note two things: (i) the observed
N(\ion{C}{iv})/N(\ion{O}{vi}) ratio is not consistent with that derived
for models that involve turbulent mixing layers, and (ii)
the H-$\alpha$ emission from S147 has been reported
by Kirshner \& Arnold (\cite{kirsh79}) at a similar
velocity to that of the \ion{O}{vi} absorption.

Clearly the gas surrounding S147 has been highly
disrupted by the initial SN explosion, and thus it is of
great interest to investigate how the passage of fast-moving
interstellar shocks and the influence of 
local ionization processes on the ambient interstellar gas and dust
may have altered their gas-phase element abundances. 
Previous studies of high-velocity gas clouds associated with SNRs
have revealed an increase in the relative abundances of
Fe, Si and Al compared with their cosmic values, while
elements such as N, O and Ar have shown an opposite
behavior (Savage \& Sembach \cite{savage96}; Jenkins
et al. \cite{jenkins84}; Welsh et al. \cite{welsh01}).

The interstellar abundance of a given element, X, relative to that
of hydrogen (H) is normally defined as [N(X)/N(H)]$_{cloud}$,
where N(X) and N(H) are the total column densities of
the elements X and H present in the gas cloud. For many elements
(except N) the greatest column-density contribution
arises in absorption lines
of their first and second ionization
states, which are observable at ultraviolet wavelengths.
These observed values can then be compared to their solar (cosmic)
abundance values (as listed in Table 5 of Savage \&
Sembach \cite{savage96}), such that an elemental depletion
or over-abundance can be determined for the gas cloud under investigation.  
Unfortunately direct measurement of the hydrogen column density for
the IV clouds observed towards both sightlines 
associated with S147 are unavailable. This is also
the case for the 
element sulfur which, because
it is normally assumed to be undepleted in the general ISM, has been
widely used as a surrogate abundance comparison element. 
We have therefore decided to use comparisons of the ratios
of the actual measured element column densities of the IV clouds
(listed in Tables \ref{table:2} and \ref{table:3}) as a potential probe of 
the variation of element abundance. We have already discussed
how the variation of the N(\ion{Na}{i})/N(\ion{Ca}{ii}) cloud column-density
ratio can be used as an interstellar gas diagnostic in Section
4.3, and the column-density comparison method for
UV absorption lines has also been
used successfully by Jenkins et al. (\cite{jenkins84})
and Trapero et al. (\cite{trapero96}) for
high-velocity gas clouds
in the ISM.

\subsection{Element abundances in the IV gas}
\begin{table*}[htbp]
\caption{Element column-density ratios in SNR 
intermediate-velocity gas compared with cold, diffuse ISM gas and cosmic
abundances}
\label{table:4}
\centering
\begin{tabular}{lccccccc}
\hline
\hline
Element Ratio&Ionization Stages&S147&S147&Monoceros Loop&Vela SNR& $\zeta$ Oph& Cosmic Ratio \\
&(SNRs)&HD36\,665&HD37\,318&&& & \\
\hline
Mg/Fe& (\ion{Mg}{i} + \ion{Mg}{ii})/\ion{Fe}{ii}& 0.88 & 0.5 & $>$1.92 & 0.71 & 35.7 & 1.18 \\
Al/Si& \ion{Al}{ii}/\ion{Si}{ii}& 0.045& 0.15& 0.028& 0.071& 0.007& 0.085 \\
Si/Fe& \ion{Si}{ii}/\ion{Fe}{ii}& 1.91& 1.07& 2.63& 2.24& 3.5& 0.72 \\
N/Fe& (\ion{N}{i} + \ion{N}{ii})/\ion{Fe}{ii}& 6.1& 3.88& 2.79& N/A& 575& 2.88 \\
O/Fe& \ion{O}{i}/\ion{Fe}{ii}&10.7& $<$1.09& 12.9& 11.2& 2240& 22.9 \\
Na/Ca& \ion{Na}{i}/\ion{Ca}{ii}&0.085  &N/A &1.14 &N/A & 166& 0.93 \\
\hline
\hline
\end{tabular}
\end{table*}

In the following analysis we use the $\it total$ column-density
value for the summed V = -65 and -56 km~s$^{-1}$ components for
the various lines of each element detected
towards \object{HD 36\,665}, as listed in Table \ref{table:3}.  For Na and Ca, we use the
column-density values for the -65 km~s$^{-1}$ component, as listed in Table \ref{table:2}.
For the \object{HD 37\,318} sight-line
we use the corresponding V = +92 km~s$^{-1}$
component column-density values.
We have also supplemented these \ion{N}{i}, \ion{N}{ii}, \ion{O}{i}, \ion{Fe}{ii}, \ion{Na}{i} and \ion{Ca}{ii} data with the
column-density values for \ion{Mg}{i}, \ion{Mg}{ii}, \ion{Al}{ii}, and \ion{Si}{ii} listed by PGB81 and PG83 
to derive the ratios of [Mg/Fe], [Al/Si],
[Si/Fe], [N/Fe], [O/Fe] and [Na/Ca]
listed in Table \ref{table:4} for 
the intermediate-velocity components observed towards
\object{HD 36\,665} and \object{HD 37\,318}.   
Note that these ratios include only the ionization stages mentioned
above.  If ionization is important, a significant amount of each element may
be in higher ionization stages.
For comparison purposes we also
present values of these element column-density ratios
for fast-moving interstellar 
gas clouds observed at V = +65 km~s$^{-1}$ towards
HD 47240 in the
Monoceros Loop SNR (Welsh et al. 
\cite{welsh01}), at V = -58 km~s$^{-1}$ towards
HD 72089 in the Vela
SNR (Jenkins et al. \cite{jenkins98}, Jenkins \& Wallerstein
\cite{jenkins96}) and
in the cold and diffuse gas at V = -15 km~s$^{-1}$
observed towards $\zeta$ Oph by
Savage et al. (\cite{savage92}) and Morton
(\cite{morton75}). All these
ratios are compared with the ratios of
their canonical cosmic
abundances (relative to that of the Sun)
as listed by Savage \& Sembach (\cite{savage96}).  

Inspection of Table \ref{table:4} reveals that the column-density
ratios obtained for IV gas in SNRs are generally quite different
from the values obtained for cold and diffuse gas of
the ISM as
observed towards $\zeta$ Oph. Clearly the physical
and chemical conditions in the disrupted gas associated
with SNRs are different to those normally encountered in
slow-moving gas of the 
general ISM.
Several authors have found that
the refractory elements of Fe, Ca, Mg, Al, C and Si are
not heavily
depleted in IV interstellar gas clouds,
a result which is easily explained
by invoking interstellar dust grain destruction 
mechanisms (such as thermal sputtering and
grain-grain collisions) that operate within 
fast-moving gas clouds and
return these elements back into the gas phase (Barlow \cite{barlow78}).   
Therefore we might expect the
[Mg/Fe] and [Al/Si] ratios in disrupted
gas to have column-density ratios similar to their
respective solar abundance ratio
values of 1.18 and 0.085. For the 4 SNR sight-lines
sampled, the measured
values suggest that this is indeed
the case. We note, however, that 
as mentioned in Section 4.3 the \ion{Na}{i}/\ion{Ca}{ii} ratio can be
also affected by local ionization processes, and our UV absorption
data taken towards \object{HD 36\,665} in S147 have shown that the IV
component is composed of both neutral and ionized gas. The
doppler $\it b$-values derived from the \ion{Na}{i} and \ion{Ca}{ii}
intermediate-velocity components
for this sight-line (listed in Table \ref{table:2}) also suggest the
presence of high-temperature
or highly turbulent gas in this particular IV cloud.
Therefore, grain destruction by fast-moving shocks may
not be the sole process responsible for the level of
depletion of Ca revealed by our data.

We note that the [N/Fe] ratios for both S147 sight-lines are a factor of 1.3 to
2.1 larger than the cosmic ratio, whereas [O/Fe] is smaller by a factor of 2 to
20.  We have inspected the $\it FUSE$ spectral data
in the low S/N channels of SiC2a and SiC1b recorded towards both
\object{HD 36\,665} and \object{HD 37\,318} and have found a strong absorption feature centered 
at 989.8\AA\ that extends over a velocity range of $\pm$ 100 km~s$^{-1}$
in both sight-lines. Inspection of the atomic line list
by Morton (\cite{morton03}) reveals the presence of the strong
ground-state line of \ion{N}{iii} at 989.799\AA. Unfortunately this line
is blended with the strong interstellar line of \ion{Si}{ii} at 989.87\AA\
and hence it is impossible for us to assess the column density
contribution of \ion{N}{iii} in either sight-line. However, it is clear that
the $\it total$ column density of N(\ion{N}{i} + \ion{N}{ii} +\ion{N}{iii}) can only
be higher than the one used in deriving the observed [N/Fe] ratios.  In
addition, the N lines may be saturated, which would create a tendency to
{\it underestimate} the N column density.
Thus these ratios for the S147 sight-lines in Table \ref{table:4} 
must really represent lower limit values.
The apparent deficiency of O with respect to Fe can be
attributed to ionization effects, in that
radiation from shock fronts in these disturbed regions photo-ionizes
atoms to higher ion states that are
generally unobservable (Jenkins et al.
\cite{jenkins98}). 
For the case of O, the inability to observe
the contribution of interstellar \ion{O}{ii} to the total sight-line
column density (of neutral and ionized gas) has been estimated
from standard photo-ionization models
to be up to $\sim$ 0.2 dex by Cardelli et al.
(\cite{cardelli95}) and Sembach \& Savage (\cite{sembach96}).
Thus, this missing \ion{O}{ii} will increase the ratio values
for [O/Fe] listed in Table \ref{table:4}, such that the
the overall depletion level of oxygen towards S147 is
probably small. Thus in summary, although
the lack of knowledge of the column density of hydrogen
in the IV gas clouds seen towards S147 precludes us from deriving
absolute values for element depletion and abundances, our
measured column-density ratios suggest that 
the refractory elements (such as Fe, Si and Mg) are present in near-solar
abundance ratios, while N and O are not heavily depleted, and that N
may in fact be slightly overabundant.
Such results are
consistent with abundance studies of gas associated with
other SNRs. 

\section{Conclusion}
We have obtained high-resolution
(3 km~s$^{-1}$) spectra of the interstellar
\ion{Na}{i} and \ion{Ca}{ii} absorption lines observed towards 3 stars with
distances raging from 360  - 1380pc in sight-lines towards
the Shajn 147 SNR.
The two most distant stars, \object{HD 36\,665} (d = 880pc)
and \object{HD 37\,318} (d = 1380pc), lie beyond the SNR gas and
possess complex absorption profiles
in which components have been 
detected over a velocity range of -65 to
+ 80 km~s$^{-1}$. The foreground star, \object{HD 37\,367}, exhibits
none of these intermediate velocity (IV) components.   
We have fit all the absorption components observed in the
\ion{Na}{i} and \ion{Ca}{ii} lines towards
the 3 stars with models of cloud-component velocity,
doppler broadening and column density. For the well
resolved IV components at V$_{helio}$ = -65 and +39 km~s$^{-1}$ observed
towards \object{HD 36\,665} we have obtained N(\ion{Na}{i})/N(\ion{Ca}{ii}) column-density
ratios of 0.085 and 0.054 respectively. Such low ratio values
can be best explained by the influence of shocks and ionization
on the disturbed SNR gas clouds.

We have also obtained far-UV absorption spectra along the sight-lines
to \object{HD 36\,665} and \object{HD 37\,318} using the NASA $\it FUSE$ satellite.
These observations have revealed IV features at
velocities of -65 and -52 km~s$^{-1}$
in the interstellar absorption lines of \ion{N}{i}, \ion{N}{ii}, \ion{O}{i} and
\ion{Fe}{ii} seen towards \object{HD 36\,665}. Based on the detection of \ion{Na}{i}
and \ion{Ca}{ii} components at these velocities, this IV cloud is
probably
composed of both neutral and ionized gas shells.  
For the case of the 1380pc sight-line to \object{HD 37\,318}, IV 
gas was only detected in the UV lines of
\ion{Fe}{ii} and \ion{N}{ii} at 
V = +90 km~s$^{-1}$, suggesting that 
this outwardly moving cloud is composed mainly
of warm and ionized gas. The inability to detect
IV components with negative velocities towards this
star (which resides beyond the Shajn 147 SNR), suggests
that the velocity structure of the expansion of the SNR
is not coherent and that randomly
distributed large-scale inhomogeneities
must exist throughout the ambient interstellar clouds.

Highly ionized gas has been 
detected at V $\sim$ +40km~s$^{-1}$ in the lines of \ion{O}{vi}$\lambda$ 1032\AA\
along the sight-lines to both stars. The column-density ratio
of N(\ion{C}{iv})/N(\ion{O}{vi}) $<$ 0.27 is consistent with 
that predicted for
an evolved supernova by Slavin \& Cox (\cite{slavin92}).
However, this ratio is also similar to that observed
in many sight-lines passing through hot gas in the general ISM
(Spitzer \cite{spitzer96}). Therefore, we are unable to definitely
associate this high-ion absorption with gas in Shajn 147. 

We have derived estimates of the observed column-density ratios
of [Mg/Fe], [Al/Si], [Si/Fe], [N/Fe], [O/Fe] and [Na/Ca] for 
the intermediate-velocity clouds
seen in the sight-lines towards Shajn 147 and
also towards fast-moving gas in the Monoceros Loop and Vela SNRs.
The ratios for 
IV gas clouds in all these SNRs are generally quite different from those
values seen in the diffuse and cold ISM. Such differences are
almost certainly due to the destruction of interstellar dust
grains in these highly disturbed regions that returns elements
back into the gas phase. For the refractory elements
of Fe, Si, Al and Mg we observe a column-density ratio
pattern that is similar to that derived for undepleted (solar
abundance) gas.  
The elements of N and O are greatly influenced
by the ionization conditions in these regions,
and when the column-density
contribution from the unobserved species of \ion{N}{iii} and \ion{O}{ii}
are accounted for, it appears that neither N nor O are significantly depleted.
 
\begin{acknowledgements}
We are grateful to the staff and directorate of the Observatoire
de Haute Provence (France) for the ground-based observations.
We would also like to recognize members
of the $\it FUSE$ project team at Johns Hopkins University.
Financial support for both BYW and SS came from Guest
Investigator funding 
from the NASA $\it FUSE$ project
under contract NAS5-32985 to the Johns Hopkins University. 
\end{acknowledgements}


\end{document}